# Black Silicon with high density and high aspect ratio nanowhiskers

S Kalem [a], P Werner [b], Ö Arthursson [c], V Talalaev [b], B Nilsson [c], M Hagberg [c], H Frederiksen [c] and U Södervall [c]

[a] TUBITAK- BILGEM, National Research Institute of Electronics, Gebze, 41470, Turkey
[b] Max-Planck-Institute, Department of Experimental Physics, Halle(Saale) D-06120 Germany
[c] Department of Microtechnology and Nanosciences, Chalmers University of Technology, Göteborg, Sweden

e-mail: s.kalem@uekae.tubitak.gov.tr

**Keywords:** Silicon, black silicon, silicon grass, nanowhiskers, tapered nanowires, photoluminescence, field emission diode, nanopillars, microscopy, ellipsometry

**ABSTRACT**

Physical properties of black Silicon (b-Si) formed on Si wafers by reactive ion etching in chlorine plasma are reported in an attempt to clarify the formation mechanism and the origin of the observed optical and electrical phenomena which are promising for a variety of applications. The b-Si consisting of high density and high aspect ratio sub-micron length whiskers or pillars with tip diameters of well under 3 nm exhibits strong photoluminescence (PL) both in visible and infrared, which are interpreted in conjunction with defects, confinement effects and near band-edge emission. Structural analysis indicate that the whiskers are all crystalline and encapsulated by a thin Si oxide layer. Infrared vibrational spectrum of Si-O-Si bondings in terms of transverse-optic (TO) and longitudinal-optic (LO) phonons indicates that disorder induced LO-TO optical mode coupling can be an effective tool in assessing structural quality of the b-Si. The same phonons are likely coupled to electrons in visible region PL transitions. Field emission properties of these nanoscopic features are demonstrated indicating the influence of the tip shape on the emission. Overall properties are discussed in terms of surface morphology of the nano whiskers.

## 1. Introduction

Black Silicon or Si grass has been receiving a great deal of attention due to their interesting physical properties and promising potential technological applications in the field of energy, sensing and emitting [1-4]. The b-Si was produced for the first time during the formation of Si trenches by reactive ion etching (RIE) in fluorine, bromine and chlorine plasmas [5-6]. Also, it can be formed by a maskless RIE employing $CF_4$ [7] due to auto-masking of the surface at random spots. However, detailed mechanism of the black Si formation and the origin of its properties are yet to be understood. It is widely believed that the formation of nanopillars during RIE is due to a local variation of the Si etch rate. This variation in etching rate can be caused by Si surface itself, for example inhomogeneous oxide layer or incompletely removed native oxide. The plasma tool and plasma source can be a cause of a micro-masking material e.g., by products of sputtering and re-deposition of plasma chamber electrodes and masking material [8-9].

Beside above mentioned methods, a number of groups have been able to produce b-Si surfaces. Black Si can be formed by shining femtosecond laser pulses on Si followed by thermal annealing [10]. Depositing a fine-grained natural mask using $CF_4/O_2$ plasma in combination with $Al_2O_3$ which was followed by RIE in $Cl_2/Ar$ plasma yielded a b-Si with densities of $\sim 10^2$-$10^3 \mu m^{-2}$ [11]. The b-Si surface obtained by this method exhibits a very weak and broad PL peak at around 700 nm. But, after annealing in forming gas (400$^o$C, 4 hours) and capping with hydrogenated $SiN_x$, a stronger PL peak appeared at 550 nm. The position of this peak was found to be independent of the diameter of the shape of pillars. However, no light emission was found to be attributable to Si nanopillars possibly due to insufficient passivation of surface defects [12]. Silicon nanopillars were also fabricated by deep UV lithography, a highly anisotropic Si RIE based on fluorine chemistry [13]. More controlled version of Si nanopillar arrays can be prepared using nanosphere lithography with different size of spheres [14]. On the other hand, switchable wettability of b-Si was demonstrated by applying external electric fields, thus indicating potential use of b-Si [15].

A different way of producing b-Si surfaces is based on thermal treatment. It was shown that Si nanowhiskers could be formed on untreated Si(100) using electron beam thermal annealing [16]. According to these studies, native oxide layer is undergone a thermal decomposition, which occurs through the formation of voids in the oxide layer [17-18]. The void growth occurs via the interfacial reaction between Si and $SiO_2$ resulting in SiO monomers which diffuse to the void perimeter. More recent method of the b-Si formation is based on local metal-catalyzed wet chemical etching [19-20]. Up to 40% increase in short-circuit current of solar cells could be achieved using this technique. A number of reports involve electroless metal deposition for the fabrication of well aligned single crystalline Si nanowire arrays [21]. Nevertheless, more direct formation method used a mixture $SF_6$ and $O_2$ in an inductively coupled plasma (ICP) reactor [22]. It is obvious from all the previous reports that b-Si can be formed under variety of process conditions. However, it is not clear yet whether there is a common responsible mechanism of the formation of b-Si. It is the purpose of this work to advance our knowledge in this field by providing new experimental evidence on the formation and the physical properties of resultant b-Si.

The technological motivation behind this work is to demonstrate the potential applications in sensing, energy and emitting. b-Si prepared by electroless metal deposition was shown to be a promising alternative in replacing expensive photolithographic processes to make a Si base field emitter [23]. In addition to photovoltaic applications [19, 20], the possibility of using b-Si in mechanics [22], microfluidics [25], process optimization [6], micro electromechanical systems (MEMS) [26] and terahertz emission [1] shows that the applications cover a broad range of fields. Our exploratory level research effort provides further evidence for potential applications by improving material quality, controlling the process and understanding the mechanism of the formation and observed properties.

## 2. Experimental

In our work, the black Si was formed by reactive ion etching (RIE) process of thermally oxidized and a photoresist (PMMA) coated 3-inch size p-type Si wafers with <100> and <111> crystallographic orientations. The resistivity of the wafers used for our studies were around 10 Ohm-cm. The RIE of these wafers using chlorine plasma leads to the formation of nano whiskers on all over the surface of Si wafer. As a result of this process, whole wafer surface becomes black, since the whiskers soak all the white light. Scanning electron microscopy (SEM), Transmission electron microscopy (TEM), Spectroscopic ellipsometry (SE), Fourier transformed infrared spectroscopy (FTIR), electron dispersive spectroscopy (EDS) and photoluminescence (PL) measurements were carried out in order to investigate physical properties of b-Si surface. SEM enables to distinguish the surface morphology between the b-Si formed on Si(100) and Si(111) wafers as shown in Fig.1. TEM bright field analysis indicates the presence of tapered crystalline Si nanowhiskers with sizes well under 3 nm at tips and with lengths of up to about 500 nm (see TEM image in Fig. 2). It is possible to control their length and shape by the process time although this correlation was not explored fully. Electric field emission (EFE) diodes were fabricated on Si whiskers formed on p-Si(111) with circular contacts in a parallel plate configuration to provide better insight on the carrier transport mechanism. The electrical contacts consisted of an evaporated 20 nm Au front electrode and a 200 nm sputtered Al back contact. The active device area and the distance between electrodes are 0.03 cm$^2$ and 200 μm, respectively. The device was fabricated on p-Si wafer of 10 Ohm-cm resistivity.

To prepare b-Si surface, a thermal oxide of 200 nm was first grown on p-type 3-inch size Si(111) and Si(100) wafers (10 Ohm-cm) which was followed by PMMA thin capping layer of 250 nm (novolac-type negative e-beam resist). The wafers were then exposed to plasma generated by an inductively coupled plasma/reactive ion etching (ICP/RIE) system from Oxford Instruments using chlorine (Cl$_2$) gas. This process used a Cl$_2$ flow rate of 50 sccm and process pressure of 7 mTorr under electrode power and inductively coupled power of 50 Watts and 100 Watts, respectively. The electrode is water cooled at 17 C constant temperature and the temperature of the silicon surface is likely to be around 60-100 C. The heat transfer down to the water cooled electrode was assumed to be negligible.

PL has been excited by a HeCd laser of 10 mW at 325 nm (3.81 eV). The PL signal is collected by mirror optics and dispersed by a single 0.5m grating. The resulting emission is detected by a liquid nitrogen cooled Si CCD, InGaAs array and Ge photodetector at room temperature as well as at 10K. FTIR measurements were done at room temperature between 400 – 4000 cm$^{-1}$. SEM and TEM images were recorded using a JEOL JSM 6340F and a JEOL JEM 4010, respectively. The energy dispersive spectroscopy for concentration measurements was carried out using SEM, JEOL-JSM-6335S tool.

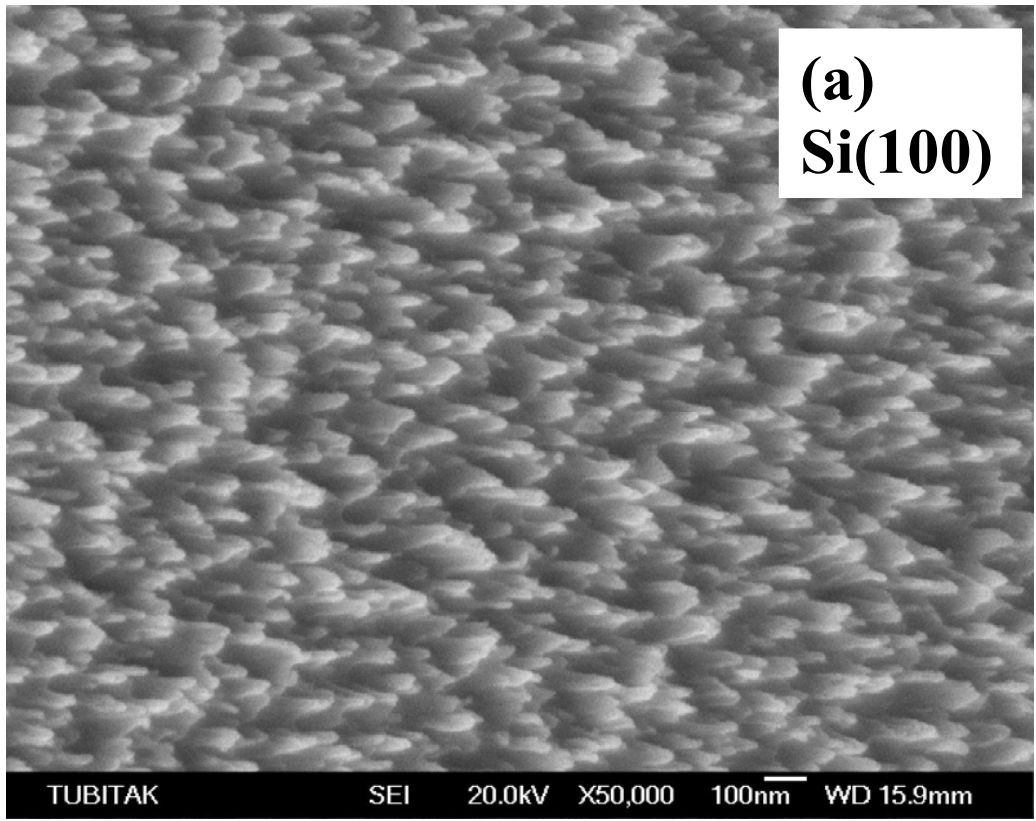

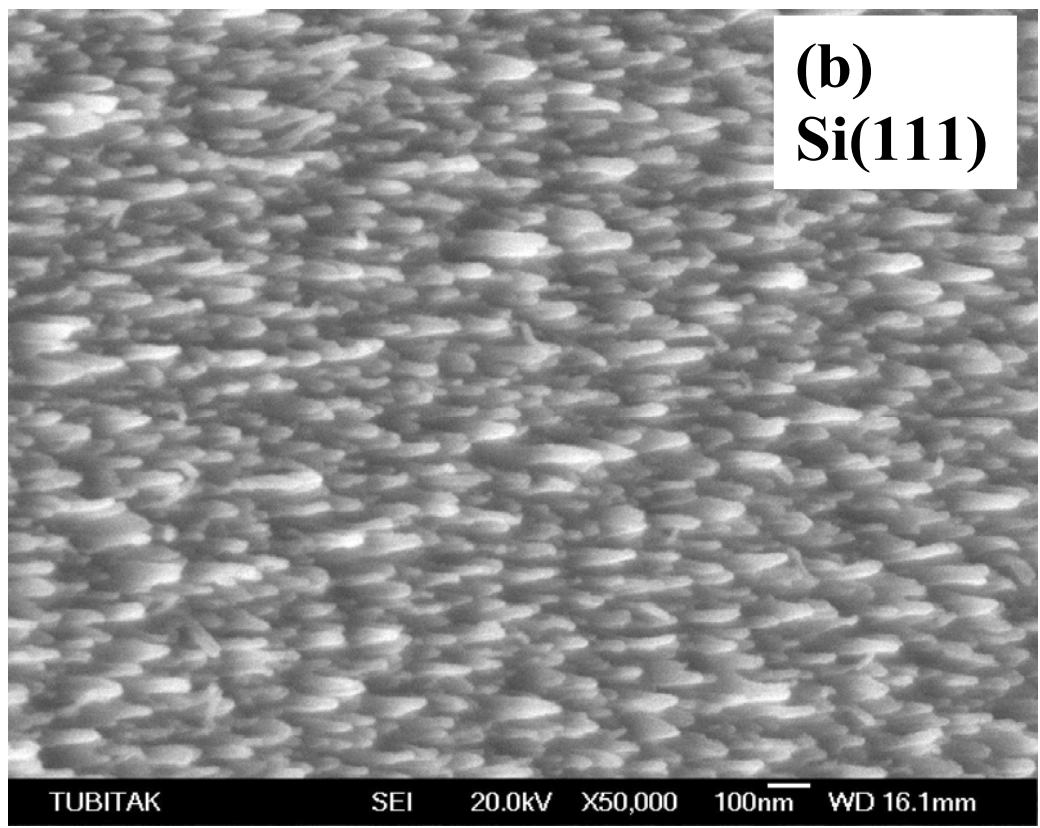

**Figure 1** SEM images of b-Si fabricated on **(a)** Si(100) and **(b)** on Si(111) by ICP/RIE using $Cl_2$. The surface images were taken at a tilt angle of $45^o$.

## 3. Results

SEM images of black Si (b-Si) consisting of random needle like nano whiskers fabricated by ICP/RIE on p-Si(100) and p-Si(111) wafers using chlorine plasma are shown in Fig. 1. The process time was 15 minutes for both of the wafers and the images are taken at 45º tilt angle. Note that the surface morphology is different for two types of wafer. The whiskers fabricated on Si(111) are longer and have regular shapes while those on Si(100) are shorter with irregular shapes representing a rougher surface. The length of the whiskers ranges from about few tens of nanometer to about 450 nm with an average length of 320 nm for those fabricated on Si(111) as shown in Fig. 1(b). Those formed on Si(100) are shorter and have an average length of 260 nm. Further detailed images are obtained by TEM analysis performed on b-Si fabricated on Si(100) indicating that Si whiskers are needle-like tapered structures pointing vertically outward from the wafer plane and are all crystalline as shown by Bragg reflections in Fig. 2(a). The whiskers are randomly distributed over the Si wafer and have a surface density of 250-550 $\mu m^{-2}$ (or 25-55 x $10^9$ $cm^{-2}$) and the aspect ratios (length to average width at half maximum) of greater than 20. Their diameter at the bottom of the whiskers ranges from 30 to 50nm. The tip diameter of the nano whiskers can be much smaller than 3 nm as shown in Fig. 2(a) and Fig. 2(b) more explicitly. The same figure clearly indicates that the whiskers are crystalline as evidenced by lattice fringes of atom planes as shown at the insert of Fig. 2(b).

The surfaces of individual nanowhiskers do not represent a smooth surface structure. Nanowhisker surfaces as shown in Fig. 2(b) contain a number of defects like missing atom planes, the presence of a number of kinks and vacancies. Also, there is an uneven encapsulation of the surface by an oxide layer of about 5 Å thick. The presence of this oxide layer or oxidation of the whiskers is also confirmed by EDS and FTIR analysis as demonstrated in the following section.

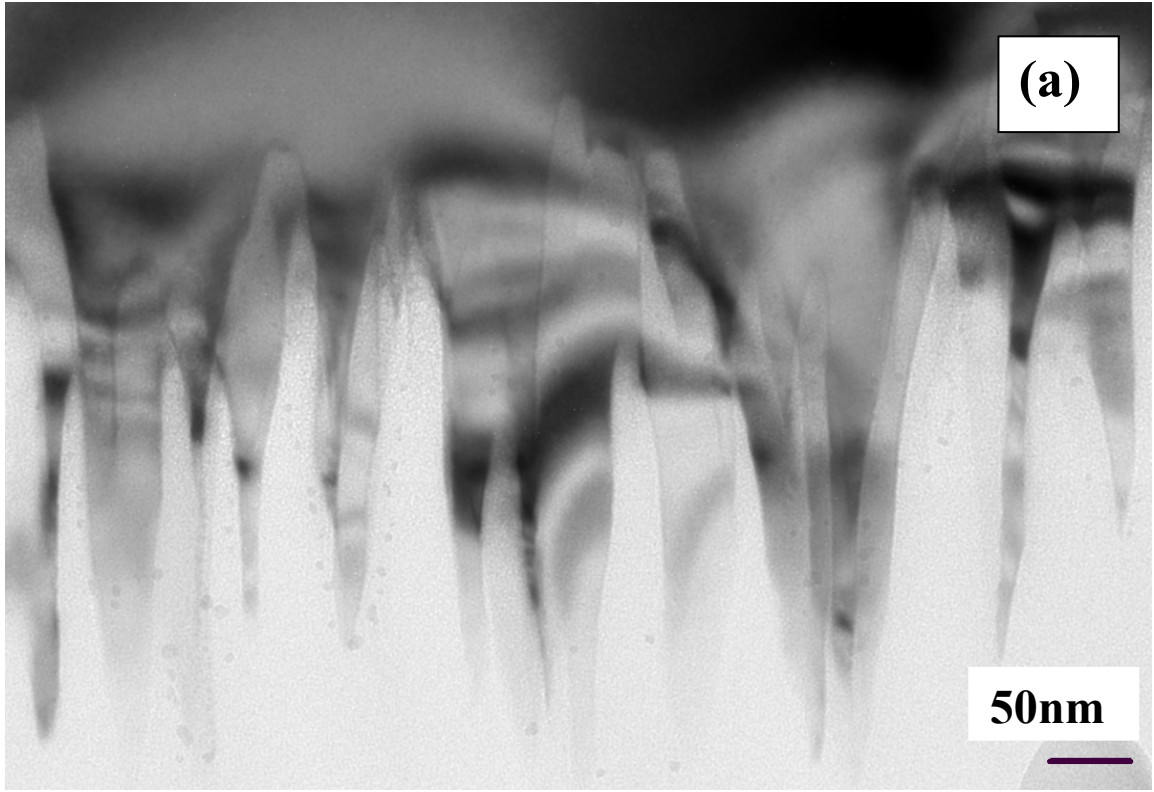

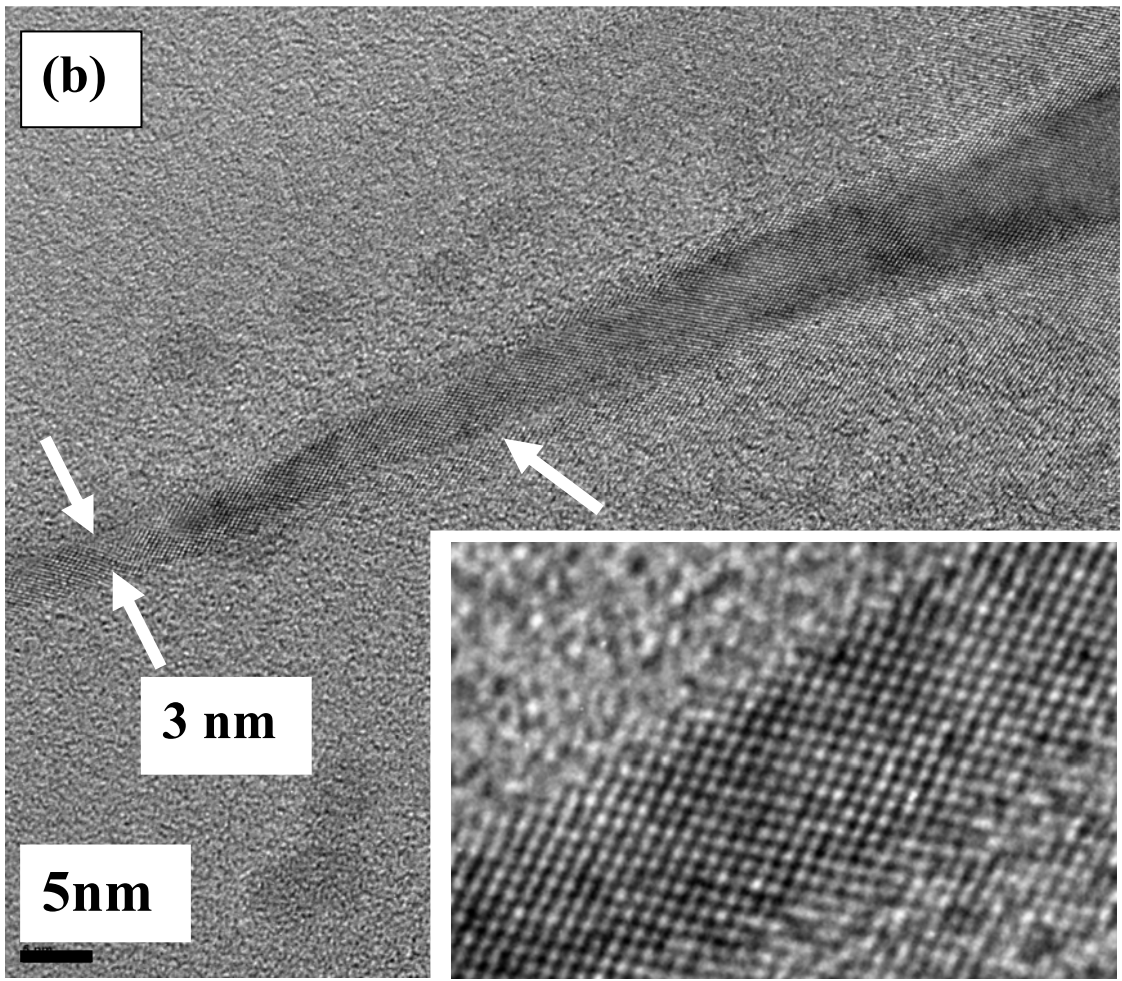

**Figure 2**. *(a) A typical TEM image of sub-micron size Si whiskers as fabricated on Si(100). (b)A single silicon whisker exhibiting crystalline structure as evidenced by Bragg reflections through atomic planes whose crystalline lattice was explicitly displayed at the insert in Fig. 2b.*

As shown at the insert in fig. 2(b), the surface of whiskers is not smooth containing missing atom planes and kinks representing a porous like structure. This structure is indicative of the presence of an ultra thin (about **5 Å**) non-crystalline (or a porous-like oxide structure) layer on the pillars. Actually, the EDS analysis supports the fact that the nanowhiskers are encapsulated by such an oxide layer. EDS probes up to 4.5 at.% oxygen confirming the presence of this oxide layer (possibly SiOx, 1.5<x<2) on the whiskers. The oxide encapsulation is also confirmed by FTIR through the presence of Si-O-Si stretching vibrations at 1085 cm$^{-1}$ as shown in Fig. 3. From previous studies correlating the Si-O frequency with x, one can estimate that 1085 cm$^{-1}$ would correspond to a sub-oxide SiOx of x=2, that is SiO$_2$ [27]. Another important feature of these results is the appearance of a second band at 1255 cm-1 when the incident beam is oblique. Fig. 3 represents the change of the spectrum for 60$^o$ angle of incidence.

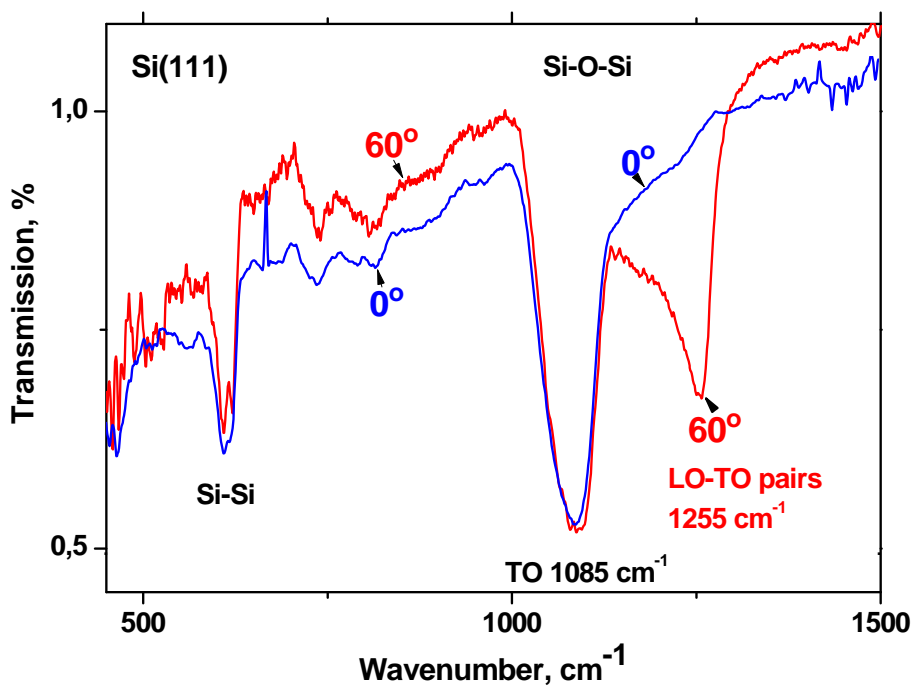

**Figure 3** *Infrared spectrum of b-Si on Si(111) wafer taken at oblique incidence(blue:0$^o$ and red:60$^o$) indicating the importance of LO-TO disorder induced mode coupling effect.*

Black Si consisting of nanowhiskers exhibits interesting PL properties at room and low temperature as shown in Fig. 4 for a b-Si produced on p-Si (111) wafer. At room temperature (RT), PL measurements reveal a broad asymmetric emission band in the visible region with a low energy shoulder and a tail at high energies as shown in Fig. 4(a). From the deconvolution of this spectrum, the peak positions of these bands are found to be at 595 nm (2.08 eV), 649 nm (1.91 eV) and 696 nm (1.78 eV). Relative intensities indicate the domination of the band at 696 nm. At low temperature (10K), the emission is a blue shifted with the RT emission components at low energy shoulder Fig. 4(b). The deconvolution of the spectrum reveals the presence of the following bands at 610 nm, 649 nm, 699 nm with a dominant emission band at 560 nm. Note that the relative intensities of the bands at 610 nm and 650 nm are almost unchanged as compared to RT emission. However, the intensity of the band at 696 nm decreases significantly. Much weaker emission intensities and lower energy bands have been reported from the black Si formed by femtosecond laser pulses [10] and those produced in $CF_4/O_2$ plasma using $Al_2O_3$ cathode [11]. These emission features have been attributed to band tail defect states or impurity mediated recombination mechanisms [10, 28]. Also, we have already shown that there is a significant contribution from the b-Si to the broad PL emission originating from a photonic structure of Si rods [29].

Our b-Si exhibits also an infrared PL near the band edge (1.09 eV) of the bulk Si which is attributable to band-to-band (BB) free carrier recombination. Fig. 4(c) compares the RT PL spectra of whiskers formed on Si(100) and Si(111) wafers. The BB emission intensity for those fabricated on Si(111) is much stronger probably due to their enhanced light harvesting efficiency due to their higher aspect ratios of ~30. Similar PL observations were reported for epitaxial Si nanowires and the BB emission was attributed to spatial confinement effect in Si nanopillars [30]. In our previous studies on photonic matrix consisting of Si nanopillars, it was shown that enhanced optical absorption at the tip region is likely followed by a fast diffusion of excitons to lower potential energy regions toward the bulk Si where the recombination takes place [29]. Tapering effect can further enhance the diffusion due to internal electric field and thermalization [31]. However, one can not rule out the theoretical findings indicating that the selection rules can be relaxed at low dimensional systems, thus enabling BB transitions at the zone center [32].

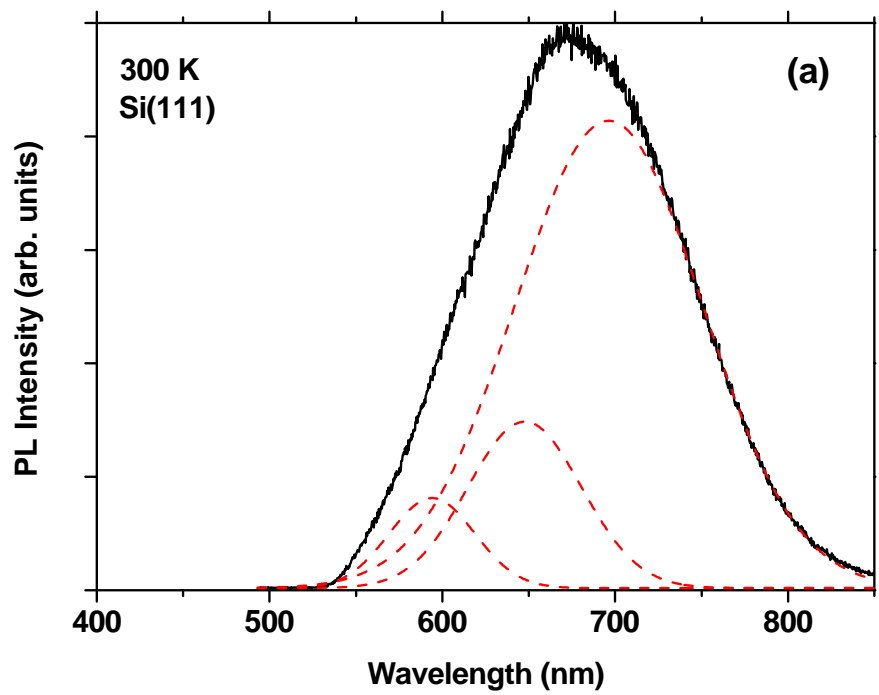

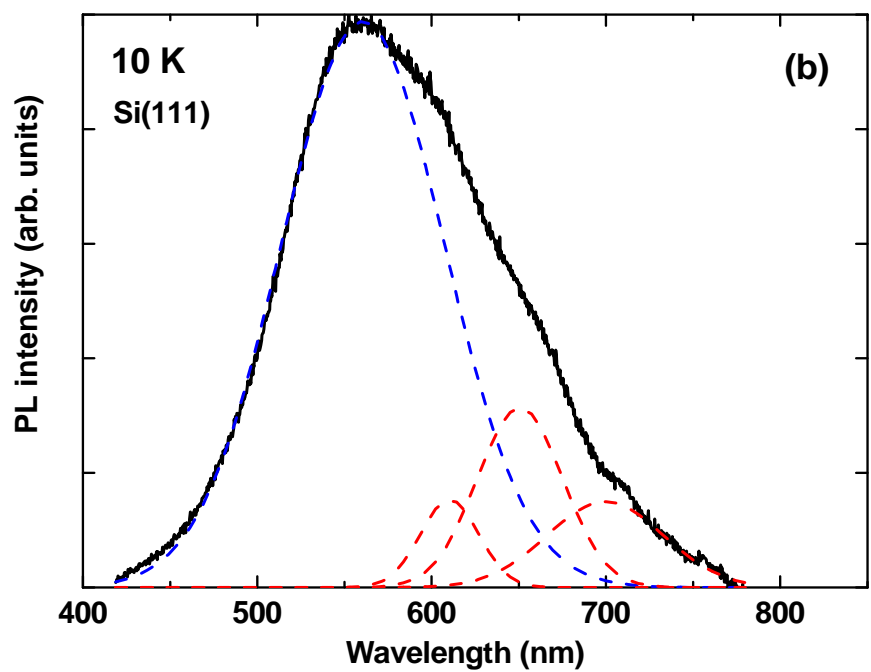

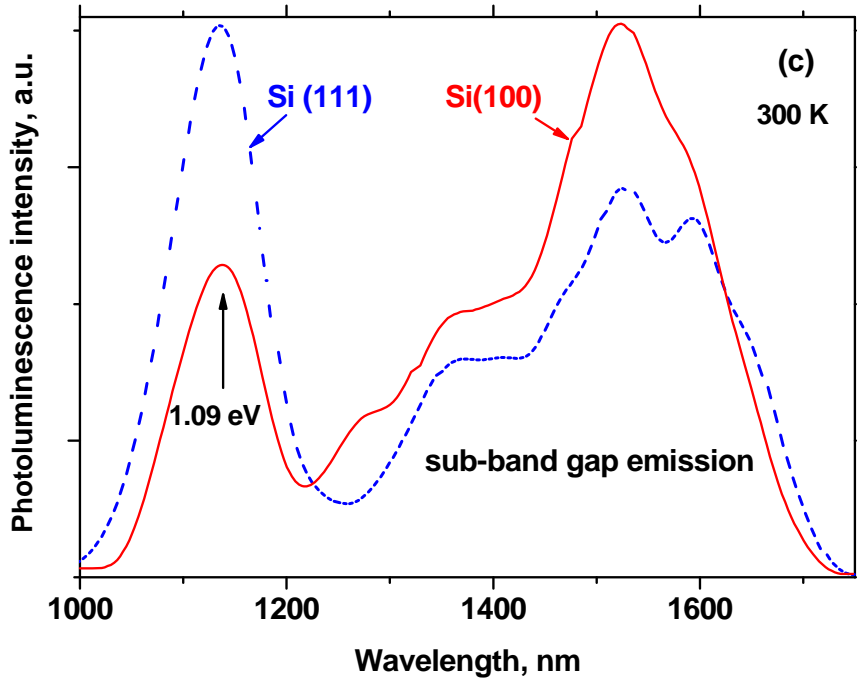

**Figure 4**. *Normalized intensity of the photoluminescence measurements obtained from black Si formed by RIE of PMMA coated p-Si(111) wafers in chlorine plasma : (a) the emission in the visible region at 300 K, (b) the emission at low temperature 10K  (c) compares the Infrared emission from black Si fabricated on p-Si(111) and p-Si(100) at 300K.. The PL excitation used a 325 nm(3.82 eV) HeCd laser line (23 mW). The results of the deconvolution are shown as dashed lines in (a) and (b).*

Fig. 4(c) shows also the emission bands originating from the sub-band gap region, which are likely due to extended defects, particularly dislocations like D1 or even impurities [33,34]. An important feature of the infrared emission is the presence of two competing recombination mechanism: band-edge and defects emission. For the b-Si on Si(100), the sub-gap emission is much stronger than the band-to-band emission, while the BB recombination dominates over the sub-gap emission in b-Si on Si(111). This is the clear demonstration of the presence of a correlation between the crystal orientation and the photoluminescence in two sets of b-Si samples. The b-Si formed on Si(111) exhibits much lower densities of deep traps than those fabricated on Si(100).

Spectroscopic ellipsometry (SE) gives clues about the electronic band structure and dielectric properties of the Si whiskers. It measures the spectral dependence of ellipsometric angles $\psi$ and $\Delta$ related to relative reflection coefficient [35]

$$\rho(E) = \tan\psi \, e^{-i\Delta}$$

The results of such measurements for angle of incidence of 75° are shown as an insert in Fig. 5 for a b-Si structure formed on Si(100). In order to deduce the critical point energies $E_1$ and $E_2$ for direct transitions near the L and X points of the Brillouin zone, respectively, the second derivative of the dielectric function $\varepsilon(\omega) = e_1(\omega) - ie_2(\omega)$ was obtained. The SE data were smoothed using the Savitzky-Golay routine in order to minimize noise. As shown in Fig. 5, the critical points $E_1$ at ~ 3.40 eV and $E_2$ at ~ 4.40 eV are very strong and their position shows a slight dependence on the whisker size. Their energetic positions are comparable with those of bulk Si [36] with slightly higher (~3%) energy for $E_2$. The relative strength is likely due to an enhancement in optical absorption in whiskers. Fig. 5 compares the second derivative of the dielectric function for the whiskers grown on Si(100) and Si(111) where the $E_1$ and $E_2$ bands for whiskers in Si(100) are slightly blue shifted by 28 meV and 82 meV, respectively. These shifts can be attributable to the fact that the whiskers grown on Si(100) are much shorter than those grown on Si(111). One possible explanation is that shorter whiskers can lead to further quantization due to further confinement effect. The whiskers on Si (100) are 20% shorter (average length is about 260 nm) than those on Si(111) (320 nm) as estimated from SEM micrographs. Indeed, it has been shown that the absorption edge can be blue shifted with decreasing Si wire length [37]. Relative blue-shift in $E_2$ peak as compared to bulk Si is supportive of these considerations. However, no blue-shift was observed in the BB transitions as shown in Fig 4b. It is possible that other effects such as the presence of strain in whiskers could cause such shifts in critical point energies [38].

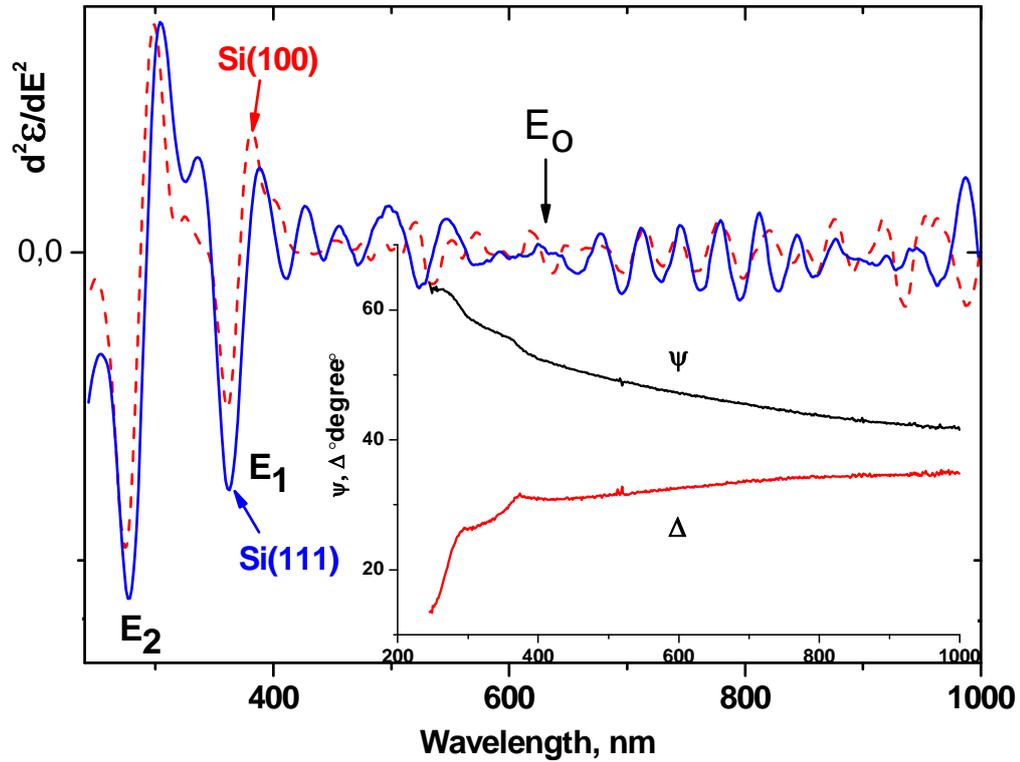

**Figure 5.** *Second derivative of the dielectric function ε indicating the observed critical points in the electronic band structure for b-Si fabricated on p-type Si(100) and Si(111) wafers. Raw spectroscopic ellipsometry data ψ and Δ variables measured at an angle of incidence of 75° for b-Si grown on Si(100) are shown at the insert.*

## 4. Discussion

The series of oscillations between 650-1000 nm in Fig. 5 correspond to interference fringes due to multiple reflections caused by an effective nanowhisker and interface layer. However, where the absorption in the b-Si structure starts, the amplitude of the oscillations damps out. The damping of the fringes near 650 nm (1.91 eV) can directly be related to a strong optical absorption edge [39]. By this damping of interference fringes, one can deduce the first strong direct energy gap $E_o$. This energy actually corresponds well to the observed room temperature PL peak in our samples (see Fig.4a). The major E1 and E2 peaks near 3.4 eV and 4.4 eV are due to direct transistions near L(Γ) and X points of the Brillouin zone [36].

The optical absorption coefficient at these energies is high, that is $10^6$ cm$^{-1}$, corresponding to a penetration depth of ~150 Å. This value confirms that SE probes the dielectric properties of Si nanowhiskers. Strong oscillator strengths can be attributed to an enhanced optical absorption in low dimensional Si structure.

The analysis of the PL spectra taken at RT confirms the presence of three peaks at around 600 nm (2.05 eV), 650 nm (1.91 eV), 700 nm (1.77 eV) as shown in Fig. 4(a). But, the intensity of these bands is reduced significantly at low temperature. While the relative intensity change for the first two bands is the same, the band at around 700 nm significantly looses its intensity in favor of the band at 560 nm (2.22 eV). This intensity correlation between the two bands indicates a competing recombination mechanism among them. This behavior can be explained by quantum confinement of excited carriers at the tips, in SiO$_2$ nanoparticles or carrier trapping effects at defects/impurity states in SiO$_2$. At low temperatures, excited carriers are localized at these weak potential wells or traps formed by quantum wells or defects/impurities close to band edges. As the temperature is increased, they are thermally excited and diffused to deeper potential wells and thus enhancing the intensity of the peak at around 700 nm. In both cases, it is most likely that the radiative recombination is mediated by localized defect states particularly at the surface of the whiskers or in SiO$_2$ [40]. However, further studies are required to determine unambiguously the origin of these bands. If quantum confinement is effective, the band at 560 nm and 700 nm can be attributed to recombination in Si nanostructures of 2.5 nm and 3.0 nm diameter, respectively. Actually, the size of the tips as shown in TEM images satisfies this condition. However, the presence of SiOx nanoparticles in b-Si surface can as well be the reason for PL as it was recently demonstrated in single SiO$_2$ nanoparticles [41, 42]. In these studies, the PL bands observed at about the same energies were attributed to electron-phonon coupling and exciton localization effects in Si and SiO$_2$ nanoparticles. In our spectra, the energetic separation between the PL bands are 135 meV and 155 meV, corresponding to Si-O-Si LO and TO phonons observed at 1085 cm-1 and 1255 cm-1, respectively. This is indicative of a likely involvement of these phonons in PL transitions. Similar temperature dependent behavior has already been observed in Si quantum dots embedded in SiO$_2$ [43]. The band at 700 nm is usually observed in porous Si and Si nanocrystals and attributed to quantum confinement effects [44]. The band at 560 nm has been reported in a number of studies as originating from oxygen-related defect centers in SiO2 [45, 46]. From these considerations, we attribute the bands at 700 nm and 560 nm to quantum confinement effects at the tips of Si nanowhiskers and defects, respectively.

We show that coupled LO-TO vibrational modes of Si-O can be used in assessing the importance of strain and defects on the wafer. In previous section, we have shown using EDS and TEM that there is a thin Si oxide layer on Si nanowhiskers. Figure 3 supports the findings of the EDS and TEM analysis through the presence of Si-O-Si stretching vibrations at 1085 cm-1 in FTIR spectra. However, the band at 1255 cm$^{-1}$ can be only visible at oblique incidence (for example at 60°). Note that this behavior is indicative of the presence of a disorder induced coupling of oxygen modes [47]. Coupled mode frequency TO+LO between optically active Si-O asymmetric stretching mode (in-phase motion of oxygen atoms) and inactive asymmetric one (out-of phase motion) can be observed in oblique incidence, if a coupling (due to disorder) between these motions exists. This splitting is due to the Berrman effect [48] as a result of disorder induced mechanical coupling of modes. Thus the appearance and relative increase of the splitting mode at 1255 cm-1 is a direct result of this splitting and should measure the strength of the defects/disorder in the structure.

In order to determine the diode performance and validate the fabrication procedure, electric field emission (EFE) devices were fabricated on Si whiskers. The vertical alignment of the whiskers is a great advantage in avoiding screening effects induced by adjacent whiskers [49]. Room temperature current-voltage (I-V) characteristics of such a device are shown in Figure 6, indicating extremely low leakage current. Figure 6 shows typical current-voltage characteristics (a) and the corresponding Fowler-Nordheim (F-N) plot [ln (IV$^{-2}$] versus V$^{-1}$ of the same data as an insert (b) following related field emission equation [50].

$$Ln(1/V^2) = -\beta/V + const.$$

where $\beta$ is the field enhancement factor. The straight F-N characteristics suggest field emission is occurring in the diode structure with different slopes at low and high voltage regions. Two slopes F-N emission has already been reported [51] in Si lift-off structures. The cause of this phenomenon has been attributed to a thin oxide on the tip surface or to a statistical distribution of the tips. Indeed, there is a statistical variance **$\sigma^2$** of 4700 nm$^2$ and 16 nm$^2$ in length and diameters (measured as full width at half of the whisker's length), respectively. The change of slope can then be attributed to different field enhancement factor. This factor of change was found to be about 6.0 for this particular sample. Thus the field emission property can be described by the shape of nanoscopic Si whiskers and their distribution. Lower radius tips could be the dominant influence for increased $\beta$. Tip diameter dependence of the field emission has already been demonstrated in Si pillars [52].

A plateau between two slopes is a typical of a field emission from a p-type semiconductor where the field penetration creates a depletion region and causes the current to be limited by electrons and not by the transparency of the barrier [53]. The enhancement factor $\beta$ can approximately be estimated using the following phenomenological formula [54]

$$\beta = 1 + s\,(d/r_o)$$

Where **d** and $r_o$ are the anode-to-cathode spacing and the radius of curvature for whiskers, respectively. **s** is the screening effect parameters, whose range is between 0 for very dense whiskers and 1 for a single needle. Assuming densely arranged whiskers (**s** = 0.2) and $r_o$ = 10 nm which are located in a parallel plate electrical field (**d**=200μm), $\beta$ = 2000.

The importance of sidewall surface chemistry and the production of sidewall passivation layers have been demonstrated in directional etching of Si [55]. These studies point to the role of the formation of involatile materials at sidewalls particularly. Si chloride radicals like $SiCl_X$ (X=1, 2) in our plasma have high sticking coefficient on surfaces thus they reside a long time on the surface [56] . In the case of re-dissociation of these etch products, the formation of $SiO_2$ passivation layers can be favored by a plasma enhanced chemical vapor deposition process. In-situ monitoring of the surface reactions between Si wafer and chemical species in the plasma would be very useful in confirming these assumptions.

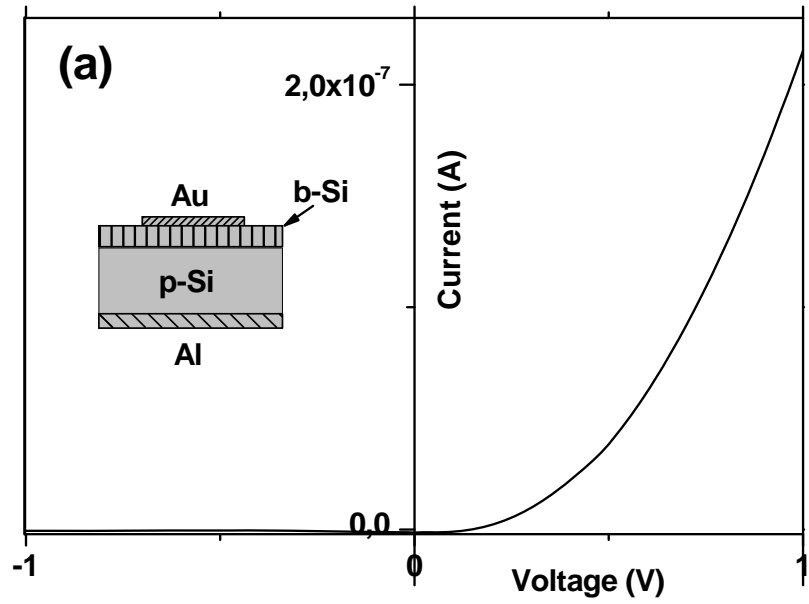

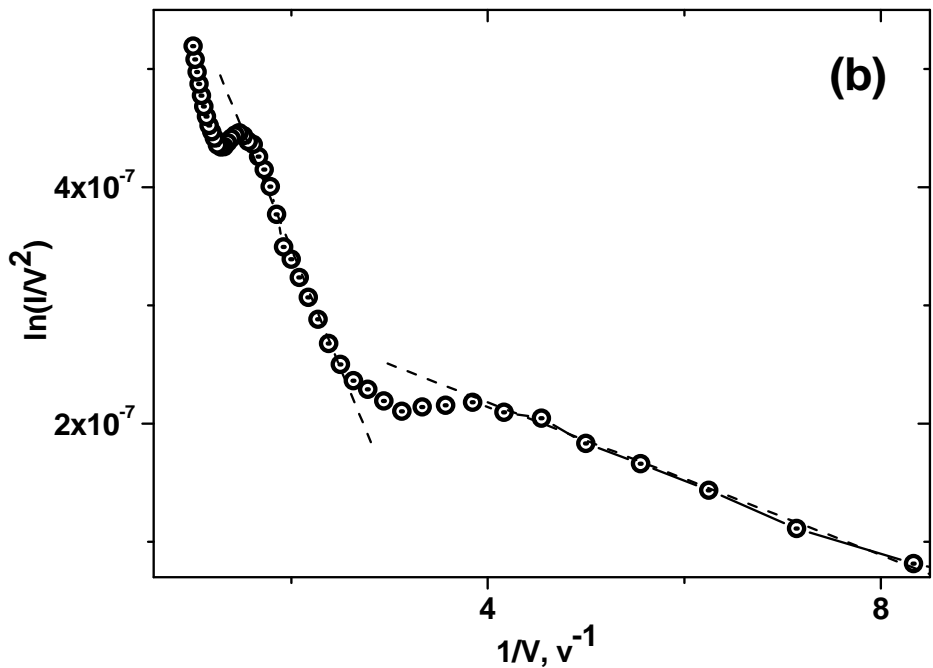

**Figure 6**. *Current-voltage characteristics of a black Silicon nanowhisker diode as measured at room temperature. (a) I-V plot and (b) Fowler-Nordheim plot. A schematic of the device structure is inserted consisting of 200 nm back contact Al and 20 nm of Au front electrode. The distance between electrodes is about 200 μm and the active device area is around 0.03 $cm^2$ as fabricated on p-Si(111) wafer of 10 Ohm-cm.*

## 5. Conclusion

Black Si consisting of high density and large aspect ratio nano-whiskers can be produced on thermal oxide grown Si wafers by ICP/RIE in $Cl_2$ plasma. Physical properties of these whiskers can be controlled by a number of experimental variables such as process time, wafer quality, and surface chemistry. Despite the absence of a supporting evidence, the condensation of Si chloride radicals plays an important role in the formation process of a black Si in our process. However, a direct experimental evidence on surface condition during the etch process needs to be generated. The role of the photoresist should be negligible since it is etched out before the oxide layer. Disorder induced optical modes (LO-TO coupling) are activated, thus suggesting that they can be used as an effective tool in assessing structural and even optical/electronic quality of the whiskers. A number of radiative recombination mechanisms exist in the emission spectrum. The b-Si is not only a dispersive element but strongly absorbing media since the oscillator strength for optical absorption is strongly enhanced at reduced dimensions where the quantum confinement is in effect. This effect results in efficient direct band transitions but also radiative transitions through the defects. Particularly, the visible emission and Si sub-band gap emissions point to the role of Si-SiO2 interfaces and extended defects, especially dislocations and impurities. Weaker sub-band gap emission and stronger band-to-band emission is indicative of a reduced deep trap density in b-Si formed on Si (111). Electron-phonon coupling in Si or $SiO_2$ (on the surface) quantum structures is likely involved in visible PL recombination mechanism. However, there is no further evidence for quantum confinement effect in the radiative recombination mechanism despite an increase in $E_2$ critical point which can be associated with strain effects as well. Instead, defect mediated emission can be more likely the dominant mechanism. Wafer orientation dependence of the defect and band-edge emission is clearly demonstrated for the infrared photoluminescence. More studies like time-resolved emission are planned to determine the origin of these interesting emission phenomena.

The possibility of fabricating crystalline Si nanowhiskers without the use of lithography offers an interesting low-cost process for novel optical and electronic component production. The black Si produced by this technique can have a significant impact on new technologies, particularly on the development of new solar cells, field emission cathodes and variety of nano-electronics devices involving sensors. Particularly, the possibility of generating light out of these structures offers promising applications in biosensing since the monolithic integration to Si circuitry would be relatively convenient. This would enable b-Si to be an attractive candidate for biological and medical applications.

**Acknowledgements**

This work was supported by BMBF-TUBITAK bilateral program under contract No:107T624, and German Federal Ministry of Education and Research (Grant No: 03Z2HN12). The authors wish to acknowledge partially funding of this work to "MC2 ACCESS" FP6 EU Program (Contr. No: 026029). Special thanks are due to Dr. Göran Alestig for fruitful discussions and his assistance in metallization and electrical measurements.

**References**

[1] Hoyer P, Theuer M, Beigang R and Kley E B 2008 Appl. Phys. Letts. **93** 091106

[2] Tokuda Y and Yagyu E 2009 Nature Photonics **3** 7.

[3] Wang Q, Page M. R., Iwaniczko E., Xu Yueqin, Roybal L, Bauer R, To B, Yuan H.-C., Duda A., Hasoon F., Yan Y. F., Levi D., Meier D., Branz H.M. and Wan T. H. 2010 Appl. Phys. Letters **96** 013507

[4] H.-C.Yuan, V.EYost, M.R Page, PStradins, D.L.Meier, H.M.Branz, *Appl. Phys. Lett.*95, 123501(2009)

[5] G.S. Oerlein, J.F. Rembetski and E.H. Payne, J. Vac. Sci. Technol. **B8**, 1199(1990)

[6] Jansen H, de Boer M, Legtenberg R and Ewenspock M 1995 J. Micromech. Microeng. **5** 115

[7] Gharghi M and Sivaththaman 2006 J. Vac. Sci. Technol. A **24** 723

[8] G.S. Oerlein, R.G.Schad and M.A. Jaso, Surface Interf. Anal. 8, 243 (1986)

[9] T.P. Chow, P. A. Maciel, and G.M. Fanelli, J. Electrochem. Soc. 134, 1281(1987).

[10] A. Serpenguzel, A. Kurt, I. Inanc, J.E. Cary, and E. Mazur, J. Nanophotonics 2, 021770 (2008).

[11] Gotza M et al. 1995 Microelectronic Engineering **27** 129

[12] Gotza M, Dutoit M and Ilegems M 1998 J. Vac. Sci. technol. **B16** 582

[13] Nassiopoulos A G, Grigoropoulos S and Papadimitriou 1996 Appl. Phys. Letts. **69** 2267

[14] Xu H, Lu N, Qi D, Gao L, Hao J, Wang Y and Chi L 2009 Microelectronic Engineering **86** 850

[15] Barberoglou M, Zorba V, Pagozidis A, Fotakis C, Stratakis E 2010 Langmuir **26** 13007

[16] S. Johnson, A.Markwitz, M.Rudolphi, H.Baumann, S.P.Oei, K.B.K.Teo and W.I.Milne, Appl. Phys. Lett. 85, 3277(2004).

[17] R. Tromp, Rubloff G W, Balk P, LeGooes F K and van Loenen E J 1985 Phys. Rev. Let. 55 2332

[18] K.Hofmann et al., Appl. Phys. Lett. 49, 1525(1986).

[19] Koynov S, Brandt M and Stutzmann M 2006 Appl. Phys. Letts. **88** 203107

[20] Koynov S, Brandt M S and Stutzmann M 2007 Phys. Stat. Sol. **1** R53

[21] Peng K Q, Wu Y, Fang H, Zhong X Y, Xu Y, Zhu J 2005 Angew. Chem. Int. Ed. **44** 2737

[22] Stubenrauch M, Fischer M, Kremin C, Stoebenau S, Albrecht A and Nagel O 2006 J. Micromech. Microeng. 16 S82

[23] Wu H-C et al 2010 J. Phys. Chem. **C114** 130

[24] Torres R et al. 2010 J. Optoelectronics and Advanced Materials 12 621

[25] Kim J and Kim C 2002 IEEE Conf. MEMS 479

[26] Lilienthal K et al 2010 Materials Science and Engin. **B 169** 78

[27] Hess P and Lambers J 2004 Microelectronic Engineering 72 201

[28] JV Anguita et al., Materials Sci.&Engin. 6, 012011(2009)

[30] S. Kalem, P. Werner, B. Nilsson, V.G. Talalaev, M. Hagberg, Ö. Arthursson, U. Södervall 2009 Nanotechnology 20 445303

[29] O. Demichel at al., Physica E 41, 963(2009).

[31] Wu Z, Neaton JB and Grossman JC 2008 Phys. Rev. B100 246804

[32] Yao D, Zhang G and Li B 2008 Nano Letters 8 4557

[33] King O and Hall DG 1994 Physical Review B 50 10661

[34] Sveinbjörnsson E.Ö. and Weber J., Thin Solid Films 294, 201(1997).

[35] Kildemo M, Hunderi O, Drevillon B 1997 J. Opt. Soc. Am. A **14** 931

[36] Pickering C 2001 Surf. Interface Anal. **31** 927

[37] Hu L and Chen G 2007 Nano Letters **7** 3249

[38] Shiri D, Kong Y, Buin A, Anantram M P 2008 IEEE 343

[39] Garriga M, Cardona M, Christensen N E, Lautenschlager, Isu T, Ploog K 1987 Phys. Rev. B **36** 3254

[40] Kanemitsu Y et al. 1993 Phys. Rev. B 48 4883

[41] Chizik A et al. 2009 Nano Letters 9 3239

[42] Martin J et al. 2008 nano Letters 8 656

[43] Wen X, Dao LV and Hannaford P

[44] Godefroo S, Hayne M, Jivanescu M, Stesmans A, Zacharias M, Lebedev O I, Van Tendeloo G and Moschalkov V V 2008 *Nat. Nanotechnol.* **3** 174

[45] Dinh LN et al. 1996 Phys. Rev. B54 5029

[46] Trukhin AN et al. 1998 J. Non-Cryst. Solids 223 114

[47] C.T. Kirk, *Phys Rev B* **38** (1988), p. 1255

[48] Berreman D W 1963 Physical Rev B **130** 2193

[49] Givargizov E I, Zhirnov V v, Stepanova A N, Rakova E V, Kiselev A N, Plekhanov P S 1995 Appl. Surf. Sci. **87-8** 24

[50] Adler EA, Bardai Z, Forman R, Goebel DM, Longo RT and Sokolich M 1991 IEEE Trans. On Electron Dev. **38** 2304

[51] JT Trujillo and CE Hunt, J. Vac. Sci. Technol. B11, 454(1993).

[52] Harvey RJ, Lee R A, Miller A J, Wigmore J K 1991 IEEE Trans. Electron Dev. **38** 2323

[53] DK Schroder et al., IEEE Trans. ED-21, 785(1974).

[54] Y-F Tzeng et al., Applied Mat. & Interfaces 2, 331 (2010)

[55] Oehrlein GS and Kurogi Y 1998 Mat Sci & Engin 24 153

[56] Cunge G et al. 2010 Plasma Sources Sci. & Technol. 19, 034017

**FIGURE CAPTIONS**

**Figure 1** SEM images of b-Si fabricated on **(a)** Si(100) and **(b)** on Si(111) by ICP/RIE using $Cl_2$. The surface images were taken at a tilt angle of $45^o$.

**Figure 2**. (a) A typical TEM image of sub-micron size Si whiskers as fabricated on Si(100). (b)A single silicon whisker exhibiting crystalline structure as evidenced by Bragg reflections through atomic planes whose crystalline lattice was explicitly displayed at the insert in Fig. 2b.

**Figure 3** Infrared spectrum of b-Si on Si(111) wafer taken at oblique incidence(blue:$0^o$ and red:$60^o$) indicating the importance of LO-TO disorder induced mode coupling effect.

**Figure 4**. Normalized intensity of the photoluminescence measurements obtained from black Si formed by RIE of PMMA coated p-Si(111) wafers in chlorine plasma : (a) the emission in the visible region at 300 K, (b) the emission at low temperature 10K    (c) compares the Infrared emission from black Si fabricated on p-Si(111) and p-Si(100) at 300K.. The PL excitation used a 325 nm(3.82 eV) HeCd laser line (23 mW). The results of the deconvolution are shown as dashed lines in (a) and (b).

**Figure 5**. Second derivative of the dielectric function $\varepsilon$ indicating the observed critical points in the electronic band structure for b-Si fabricated on p-type Si(100) and Si(111) wafers. Raw spectroscopic ellipsometry data $\psi$ and $\Delta$ variables measured at an angle of incidence of $75^o$ for b-Si grown on Si(100) are shown at the insert.

**Figure 6**. Current-voltage characteristics of a black Silicon nanowhisker diode as measured at room temperature. (a) I-V plot and (b) Fowler-Nordheim plot. A schematic of the device structure is inserted consisting of 200 nm back contact Al and 20 nm of Au front electrode. The distance between electrodes is about 200 μm and the active device area is around 0.03 $cm^2$ as fabricated on p-Si(111) wafer of 10 Ohm-cm.